\documentclass[runningheads]{llncs}
\usepackage{graphicx}
\usepackage{amsmath,amssymb,amsfonts}
\usepackage{algorithmic}
\usepackage{bm}
\usepackage{diagbox,ragged2e}
\begin{document}
\title{Propagation-aware Social Recommendation by Transfer Learning}
\author{Haodong Chang \inst{1}\orcidID{0000-0002-5015-1793} \and
Yabo Chu\inst{2}\orcidID{0000-0002-1694-9179}}
\authorrunning{H. Chang et al.}
\institute{University of Technology Sydney, Australia 
\email{haodong.chang@student.uts.edu.au}\\
\and
Northeastern University, China
\email{cyb980430@gmail.com}}
\maketitle (Haodong Chang and Yabo Chu contributed equally to this work)              % typeset the header of the contribution
\begin{abstract}
Social-aware recommendation approaches have been recognized as 
an effective way to solve the data sparsity issue of traditional recommender 
systems. The assumption behind is that the knowledge in social user-user 
connections can be shared and transferred to the domain of user-item interactions, 
whereby to help learn user preferences. However, most existing approaches merely 
adopt the first-order connections among users during transfer learning, 
ignoring those connections in higher orders. We argue that better recommendation 
performance can also benefit from high-order social relations. In this paper, 
we propose a novel Propagation-aware Transfer Learning Network (PTLN) based 
on the propagation of social relations. We aim to better mine the sharing 
knowledge hidden in social networks and thus further improve recommendation 
performance. Specifically, we explore social influence in two aspects: 
(a) higher-order friends have been taken into consideration by order bias; 
(b) different friends in the same order will have distinct importance for 
recommendation by an attention mechanism. Besides, we design a novel 
regularization to bridge the gap between social relations and user-item interactions. 
We conduct extensive experiments on two real-world datasets and beat other counterparts in terms of ranking accuracy, especially for the 
cold-start users with few historical interactions.

\end{abstract}
\section{Introduction}
\label{sec:introduction}
Nowadays, recommender systems play an essential role in providing effective recommendations to users with items of interest. The key of success is to learn precise 
user and item embeddings, where  Collaborative Filtering (CF) is the most 
traditional method \cite{hu2008collaborative,rendle2012bpr} to learn from user historical records, such as ratings, clicks, and reviews. However, for many users, it is lack of interaction data to provide accurate recommendations. The data sparsity problem limits the performance of CF-based models.

\begin{figure}[]
    \centering
    \includegraphics[width=0.7\textwidth]{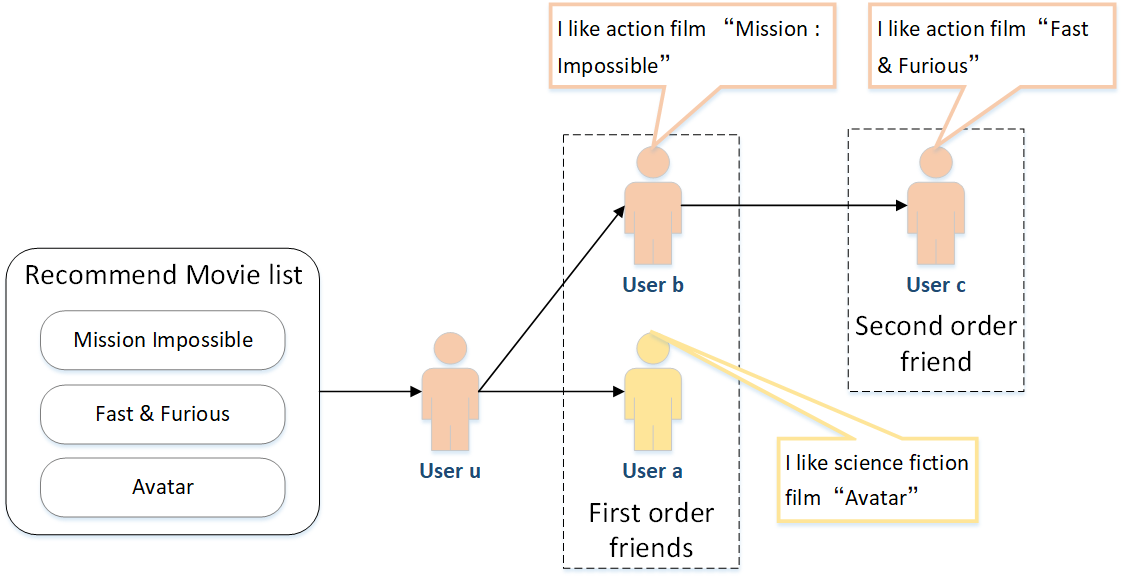}
    \caption{An example to illustrate different friends' influence. 
    User $b$ and $c$ have the same color as user $u$ indicates that they have similar preferences with user $u$, 
    and $a$ have different preferences with user $u$.
    Then the system make a recommendation by considering these friends' preference. “Mission Impossible” is in the front of the recommended list. "Avatar" is ranked behind the other two films.
    }
    \label{fig:example}
\end{figure}

With the prevalence of online social networks, 
social connections have been widely leveraged to alleviate the data sparsity problem, and formed the line of research called social-aware recommendation. 
Transfer learning \cite{xiao2017learning,chen2019efficient} is a useful approach to learn the common knowledge shared between a source domain and a target domain, and then transfer the common knowledge to enhance the model learning in target domain. Transfer learning is also applied in social-aware recommendation to learn user preference from social connections and then transfer to item domain, leading to more fine-refined user preference and thus better recommendation performance. 
%
%The common knowledge contains the information learned from the source domain, 
%which is challenging to learn from the target domain that can augment the user's representation.
%
However, most existing methods only adopt the first-order connections while ignoring the high-order connections. For example, in Figure~\ref{fig:example}, users $a$ and $b$ are both friends of user $u$ in the first order, user $c$ lies in the second order. Users $u, b, c$ share similar interests, 
while user $a$ has different interests with $u$. In this case, user $c$ (in the 2nd order) will have more positive influence on learning preference of user $u$ than first-order friend user $a$.

Therefore, we argue that high-order friends are informative and can also help learn user preference, especially considering the fact that users may not have many direct connections with other users. It is valuable to find more relevant social friends to deal with the data sparsity problem in social networks. Therefore, we adopt the trust propagation in 
our model to mine informative  knowledge hidden in high-order social relations. Specifically, social influence have been considered in two aspects. Firstly, friends in different orders will affect the learning of user preference. Different order has distinct bias towards preference learning. To the authors' best knowledge, we are the first to take into account order bias in modelling high-order social influence. Secondly, friends in the same order will have different importance for preference learning.
We apply the attention mechanism to adaptively learn the importance of friends in the same order.
Moreover, we propose a novel regularization term to 
formulate the relationship between domain-specific and cross-domain (common) knowledge to reduce the risk of model overfitting.

%Besides the mentioned above, there have some points to note 
%when mining high-order knowledge. 
%With the trust propagated in the social network, 
%each orders’ friend will influence the user preference. 
%The influence of friends is an essential source to explore 
%high-order knowledge. 
%Intuitively, the influence of each friend is different. 

To summarize, the main contributions of this paper are as follows:
\begin{itemize}
    \item We apply transfer learning to learn the sharing common knowledge between 
    social and item domains, and leverage social propagation to take into account high-order social influence for better recommendation.
    \item We propose a new factor `order bias' to distinguish social influence in high orders from low orders. We design a novel regularization term to formulate the relationship between domain-specific and cross-domain (common) knowledge and thus to avoid overfitting. 
    \item We conduct extensive experiments on two real-world datasets Ciao and Yelp, and demonstrate the effectiveness of our approach in ranking accuracy. 
\end{itemize}

\section{Related Work}
\subsubsection{Social-aware Recommendation}
Most previous social-aware recommendation works are based on homogeneity and social influence theory, 
that is, users who are connected tend to have similar behavioral preferences, and people with similar behavioral 
preferences are more likely to establish connections. The meaning reflected in the recommendation model is 
that the user's feature vector should be as close as possible to the vector space's similar user's feature 
vector. For example, \cite{zhao2014leveraging} assumed that users are more likely to have
seen items consumed by their friends, and extended BPR\cite{rendle2012bpr} by
changing the negative sampling strategy. TrustSVD\cite{guo2015trustsvd} believed that not only the user's explicit rating data and social relationships should be modeled, but the user's implicit behavior data and social relationships should also be considered. Therefore, implicit social information is introduced based on the SVD++\cite{koren2008factorization} model. Recent research has used deep neural networks as classifiers, yielding significant accuracy. E.g., SAMN\cite{chen2019social} leverages attention mechanism to model both aspect- and friend-level differences for social-aware recommendations. However, these methods use direct social connections and ignore high-order social relationships, which has a wealth of information. 

There are also some studies considering trust propagation to get high-order information. DeepInf\cite{qiu2018deepinf} models the high-order to predict the social influence. \cite{wu2019neural} proposed a DiffNet neural model with a layer-wise influence diffusion part to model how users’ trusted friends recursively influence users’ latent preferences. The further work\cite{wu2020diffnet++} jointly model the higher-order structure of the social and the interest network. However, they need to use text or image information for data enhancement, which may lack a certain degree of versatility. Moreover, existing methods ignore the influence of different order’s friends on users.

Our work differs from the above studies as the designed model uses attention mechanism to aggregate 
different friends’ influence in each order adaptively. And the influence of order are considered as order bias.
Order bias could adjust the friend's influence depend on the friend's order.

\subsubsection{Transfer Learning}
Transfer learning deals with the situation where the data obtained from different resources
are distributed differently. It assumes the existence of common knowledge structure that defines the 
domain relatedness and incorporates this structure in the learning process by discovering a shared latent 
feature space in which the data distributions across domains are close to each other.
\cite{eaton2011selective} pointed out that parts of the source domain data are inconsistent with 
the target domain observations, which may affect the construction of the model in the target domain. 
Based on that, some researchers \cite{xiao2017learning,lu2013selective} designed selective latent factor transfer models to better capture the consistency and heterogeneity across domains for recommendation. 
However, in these works, the transfer ratio needs to be properly selected through human effort and can not change dynamically in different scenarios.

There are also some studies considering the adaption issue in transfer learning. 
% However, existing methods mainly focus on task adaptation or domain adaption and neglect the useful high-order information.
\cite{cao2010adaptive} proposed to adapt the transfer-all and transfer-none schemes by estimating the similarity between a source and a target task. \cite{moon2017completely} designed a completely heterogeneous transfer learning method to determine different transferability of source knowledge.
However, these methods mainly focus on task adaptation or domain adaption. \cite{chen2019efficient} 
 propose to adapt each user’s two kinds of information (item interactions and social connections) with a finer granularity, which allows the shared knowledge of each user to be transferred in a personalized manner.
  \cite{li2020ddtcdr} propose a novel dual transfer 
 learning-based model that significantly improves recommendation performance across other 
 domains.
 Nevertheless, these methods still ignore the following two issue:1)High-order information is very helpful to improve the recommendation performance.
%  it only considers local connections while neglecting the very informative high-order information. 
 2)Sparse data in rating domain and social domain can lead to overfitting problems.

Our method innovatively leverage the high-order information for transfer learning. And we propose a novel regularization so that the user representation about the common knowledge can be reconstructed to the user representation in the social and item domains, which could reduce the risk of overfitting due to the lack of data.

% In previous studies,

\section{Our Proposed Model}

\subsection{Notations}
Suppose we have a user set $\mathcal{U}$ and an item set $\mathcal{V}$, 
let $\emph{M}$ denote the number of users 
and $\emph{N}$  denote the number of items.
Symbols $\emph{u, t}$ denote two different users, and $\emph{v}$ denotes an item. $\mathcal{F}_u$ represents the friend set of user $u$.
In social rating networks, users can form social connections with other users and interact with items, resulting in two matrices: 
user-user social matrix and user-item interaction matrix.
The user-item interaction matrix is defined as 
$\textbf{R} = \lbrack \emph{r}_{uv} \rbrack_{M \times N}$
from users' historical behaviors. $r_{uv}=1$ indicates that user $u$ has an observed interaction (purchases, clicks) with item $v$. Similarly,  we define the user-user social matrix $\textbf{X} = \lbrack \emph{x}_{ut} \rbrack_{M \times M}$ from social networks. $x_{ut} = 1$ indicates that user $u$ trusts user $t$. We represent user $u$'s embedding in three parts:  $\textbf{c}_{u}$, $\textbf{s}_{u}$ and  $\textbf{i}_{u}$, where $\textbf{c}_{u}$ denotes the latent factors shared between the item domain and social domain, i.e., the common knowledge; $\textbf{s}_{u}$ and $\textbf{i}_{u}$ are user latent factors corresponding to the social domain and item domain. The purpose of item recommendation is to generate a list of ranked items that meet user $u$'s preference.
\begin{figure*}[h]
    \centering
    \includegraphics[width=1.0\textwidth]{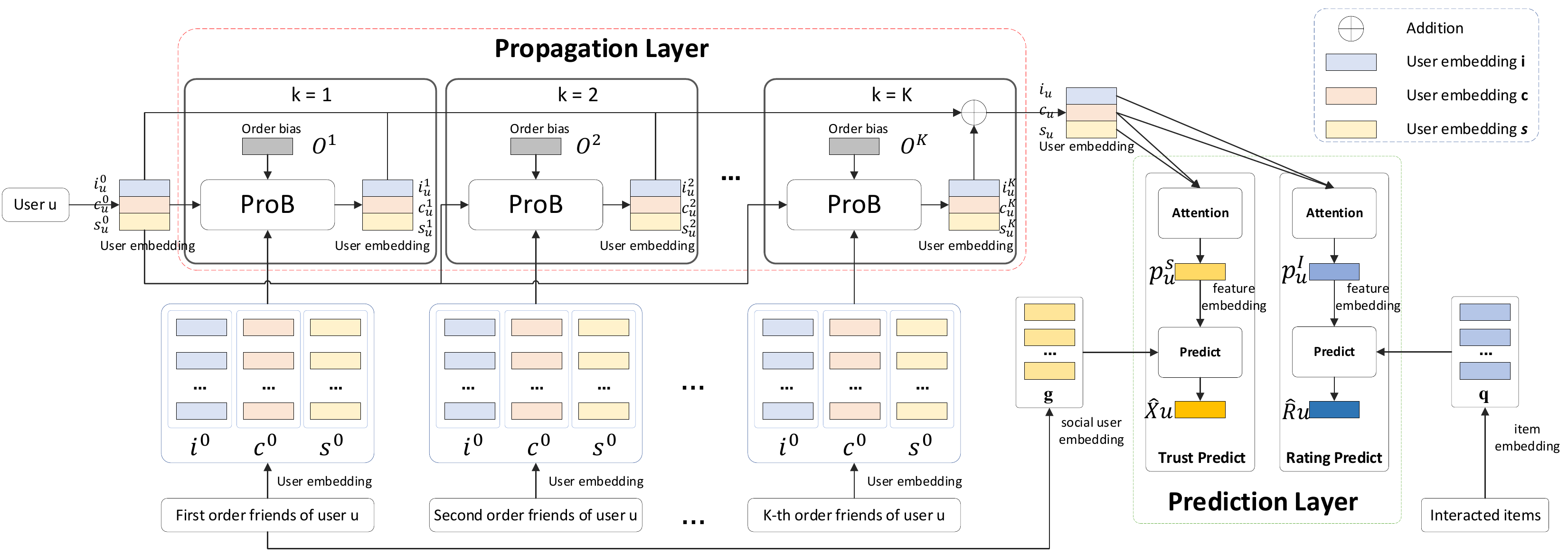}
    \caption{An overview of our PTLN model. `ProB' represents the propagation block introduced in Figure 3.}
    \label{fig:model}
\end{figure*}
\subsection{Model Overview}
The overall structure of our Propagation-aware Transfer Learning Network (PTLN) is illustrated in Figure~\ref{fig:model}. It includes three types of input: 1) the user embedding of user u and u's each order friends, 
2) the social user embedding of u's first order friends, 
3) the item embedding of the item which u has interacted. 
The outputs of our model are the predicted probability $\hat{r}_{uv}$ that how user $u$ will like item $v$, and the predicted probability $\hat{x}_{ut}$ that how user $u$ will trust another user $t$. The main architecture of PTLN contains two components: propagation layer and prediction layer.

The propagation layer propagates over social networks to incorporate the influence of high-order social friends, and then aggregate  social influence of friends in different orders. Besides, the order itself is also considered as order bias, indicating the influence bias of general friends in a specific order.  
In the prediction layer, we adopt attention mechanism to consider the domain relationships to better transfer the domain-specific knowledge and the shared knowledge for each task. Moreover, we adopt an efficient whole-data based training strategy \cite{chen2019efficient}, and involves a novel regularization term in loss function to optimize the model.

\subsection{Propagation Layer}
In this part, we aim to explore the high-order social influence based on the idea that a user may share similar preferences with her friends. As shown in Figure~\ref{fig:model}, the propagation layer are constructed in a multi-block structure. Each block's input is the user embedding of target user $u$ and that of $u$'s friends at this order.
The output is the new user embedding which includes  
high-order friends' influence. 
The new user embedding in each aspect is calculated as same in propagation block, therefore we take the process of calculating the new user embedding in common knowledge aspect as an example to explain the details of the formula. The new user embedding is learned in below four steps:

% The details of one propagation block are elaborated in Figure~\ref{fig:prob}. 
% The new user embedding is learned in four steps: 1)calculate similarity embedding between user and friends 2)calculate attention score for each friend based on similarity embedding, 3)aggregate the influence of friends based on the attention score, 4)update user embedding by friends' influence.
% The new user embedding in each aspect is calculated as same in propagation block, as show in Figure~\ref{fig:prob}. Therefore we take the process of calculating the new user embedding in common knowledge aspect as an example to explain the details of the formula.

\begin{figure}[]
    \centering
    \includegraphics[width=0.45\textwidth]{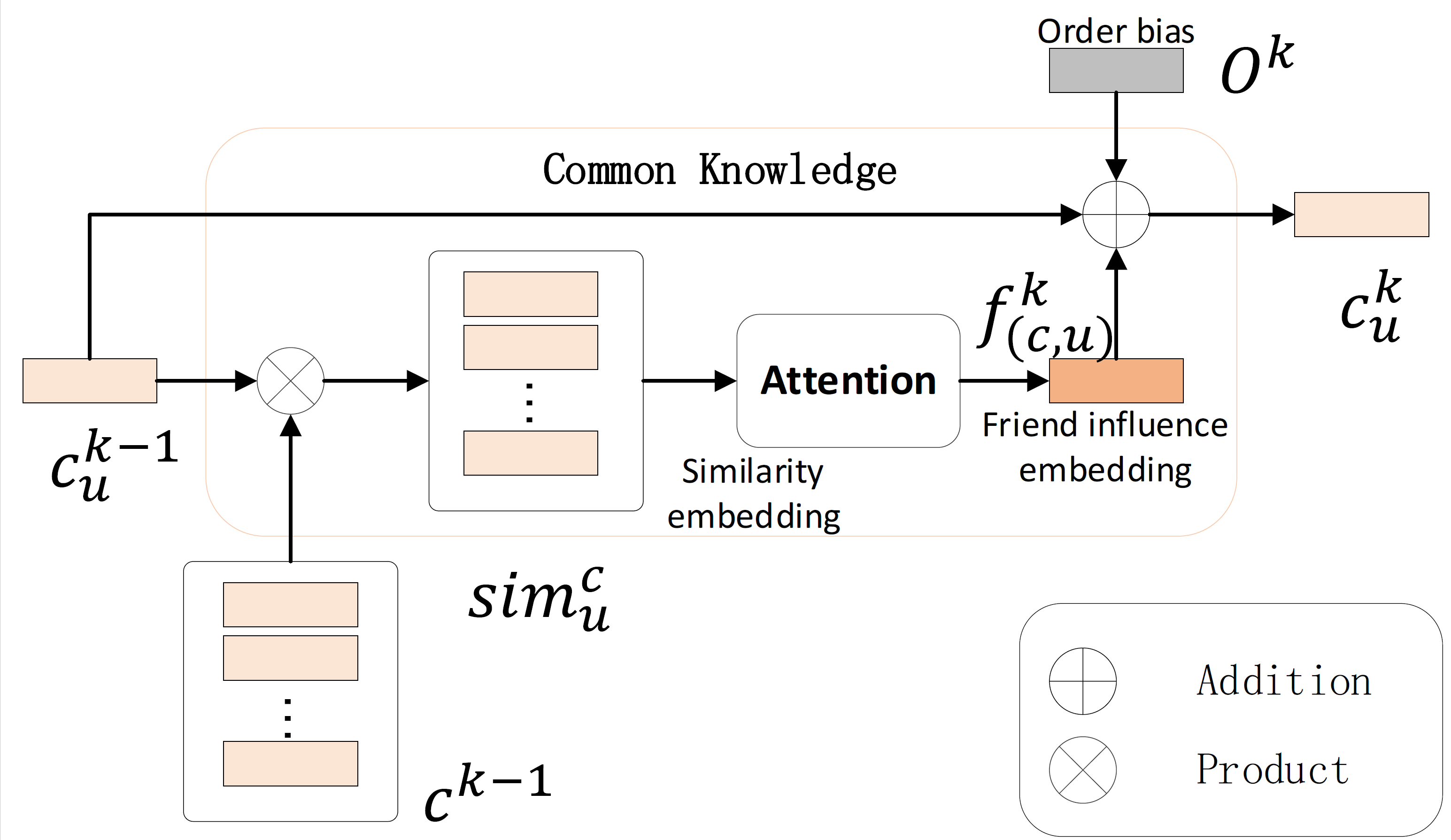}
    \caption{The details of K-th propagation block in common knowledge aspect.}
    \label{fig:prob}
\end{figure}
\textbf{1)Calculate Similarity Embedding}
% \subsubsection{1)Calculate Similarity Embedding}
User's social connection will indirectly influence the user's preference in different degrees. As discussed in the introduction, the similarity between two connected users can  be used as an essential basis for revealing the degree of influence.
Thus we adopt attention mechanism to assign the non-uniform weights 
to each friend according to the similarity between the user and her friends.
we firstly calculate the similarity embedding between user $u$ and her k-th order friend $t$ in common knowledge aspect as follow:
\begin{equation}
    \textbf{sim}_{(u,t)}^{C} = \textbf{c}_{u}^{0} \odot \textbf{c}_{t}^{0}
\end{equation}
where $\textbf{sim}_{(u,t)}^{C} \in \mathbb{R}^{D_1}$ denotes the similarity embedding 
between user $u$ and her k-th order friend $t \in \mathcal{F}_u^{k}$ in common knowledge aspect.
The superscript $0$ indicates the illustrated subject
is initial. 
$\mathcal{F}_{u}^{k}$ represents the k-th order friend set of user $u$.
the operation $\odot$ denotes the element-wise product of vectors.

\textbf{2)Calculate Attention score}
% \subsubsection{2)Calculate Attention score}
After obtaining similarity embedding from k-th order friends, the attention are calculated by a trainable weighted matrix $\textbf{W} \in \mathbb{R}^{D_1 \times 1}$.
For each aspect, the trainable weighted matrix are unique.
The k-th order friend $t$'s attention in 
common knowledge aspect $\mathcal{A}_{(u,t)}^{*(C)}$  is defined as: 
\begin{equation} 
    \mathcal{A}_{(u,t)}^{*(C)}  = \mathbf{W}_{C}^{\mathrm{\emph{T}}} \mathbf{sim}_{(u,t)}^{C}
\end{equation}
where $\mathbf{W}_{C}$ is the trainable weighted matrix to the common knowledge aspect.

Then we use the softmax function  to normalize the friend's attention score: 
\begin{equation}
    \mathcal{A}_{(u,t)}^{C} = \frac{exp(\mathcal{A}_{(u,t)}^{*(C)})}{\sum_{z \in \mathcal{F}_{u}^{k}}exp(\mathcal{A}_{(u,z)}^{*(C)})}
\end{equation} 
where $\mathcal{A}_{(u,t)}^{C}$ is the final attention of friend $t$ which indicates the degree of $t$'s influence on user $u$.

% \subsubsection{3)Aggregate Friend's Influence}
\textbf{3)Aggregate Friend's Influence}
We leverage the attention score to aggregate the k-th order friend's influence, so that the friend influence embedding we get is obtained by dynamically absorbing the influence of her friends at this order.  
\begin{equation}
    \textbf{f}_{(C,u)}^{k}=\sum_{t\in\mathcal{F}_{u}^{(k)}}\mathcal{A}_{(u,t)}^{k}\textbf{c}_{t}^{0}
    \end{equation}
where $\textbf{f}_{(C,u)}^{k} \in \mathbb{R}^{D_1}$ represents the $u$'s friend influence embedding at k-th order.

% \subsubsection{4)Update User Embedding}
\textbf{4)Update User Embedding} When generating the friend influence embedding, we merely consider the similarity between the user and friend's preference ignoring the influence of the friend's order, as discussed in the introduction. Therefore we propose a concept of order bias to model the influence bias of general friends in a specific order.
we consider that the order bias can dynamically adapt to the friend influence according to the order. 
With the friend influence embedding and order bias, the user embedding will be updated as follow:
\begin{equation}
    \textbf{c}_{u}^{k} = \textbf{c}_{u}^{0} + \textbf{f}_{(C,u)}^{k} + \textbf{o}^{k}
\end{equation}
The generated embedding $\textbf{c}_{u}^{k}$ is the new user embedding in k-th order.
$\textbf{o}_{k} \in \mathbb{R}^{D_1}$ indicates the order bias of k-th order.

After propagating with k times, we obtain k new user embedding from first order to k-th order. 
We will use all new user embedding achieved in each order with initial user embedding to generate final user embedding $\textbf{c}_{u}$ as follow:
\begin{equation}
    \textbf{c}_{u} = \sum_{k} \textbf{c}_{u}^{k}
\end{equation} 

\subsection{Prediction Layer}
% eatnn
Transfer Learning framework can transfer the shared knowledge from the source domain 
to the target domain which is a promising method of using cross-domain data to solve problems.
\cite{xiao2017learning} points that the degree of relationship between domains is varied according to the user. 
Thus, we apply the attention mechanism to use the domain-specific knowledge and common knowledge for better learning the feature embedding which represent social domain preference and item domain preference.
For a user, if the two domains are less related, the shared knowledge ($\textbf{c}$) will be penalized 
and the attention network will learn to utilize more domain-specific knowledge ($\textbf{s}$ or $\textbf{i}$) instead.
Formally, the item domain attention and the social domain attention are defined as:
\begin{small}
    \begin{equation}
            \alpha_{(C,u)}^{*} = \textbf{h}_{\alpha}^{\mathrm{\emph{T}}} \delta(\textbf{W}_{\alpha} \textbf{c}_{u} + \textbf{b}_{\alpha});
            \alpha_{(I,u)}^{*} = \textbf{h}_{\alpha}^{\mathrm{\emph{T}}} \delta(\textbf{W}_{\alpha} \textbf{i}_{u} + \textbf{b}_{\alpha}) 
    \end{equation}
    \begin{equation}
        \beta_{(C,u)}^{*}  = \textbf{h}_{\beta}^{\mathrm{\emph{T}}} \delta(\textbf{W}_{\beta} \textbf{c}_{u} + \textbf{b}_{\alpha});
            \beta_{(S,u)}^{*}  = \textbf{h}_{\beta}^{\mathrm{\emph{T}}} \delta(\textbf{W}_{\beta} \textbf{s}_{u} + \textbf{b}_{\alpha})
    \end{equation}
\end{small}
Weight matrices $\textbf{W} \in \mathbb{R}^{D_1 \times D_2}, \textbf{h} \in \mathbb{R}^{D_1}$ and bias units $\textbf{b}$
serve as parameters of the two-layer attention network. $\alpha$ and $\beta$ are related to the item domain and social domain, respectively.
$D_2$ denotes the dimension of attention network,
and $\delta$ is the nonlinear activation function $ReLU$.
    
Then, the final attention scores are normalized with a softmax function:
\begin{small}
    \begin{align}
            \alpha_{(C,u)} &= \frac{exp(\alpha_{(C,u)}^{*})}{exp(\alpha_{(C,u)}^{*}) + exp(\alpha_{(I,u)}^{*})} = 1 - \alpha_{(I,u)} ;
            \beta_{(C,u)}  &= \frac{exp(\beta_{(C,u)}^{*})}{exp(\beta_{(C,u)}^{*}) + exp(\beta_{(S,u)}^{*})} = 1 - \beta_{(S,u)} 
    \end{align}
\end{small}

$\alpha_{(C,u)}$ and $\beta_{(C,u)}$ denote the weights of common knowledge $\textbf{c}$ for item domain and social domain, respectively, which determine how much to transfer in each domain.
After obtaining the above attention weights, the feature embedding of user $u$ for the two domains are calculated as follows:
    \begin{equation}
        \textbf{p}_{u}^{I} = \alpha_{(I,u)}\textbf{i}_{u} + \alpha_{(C,u)}\textbf{c}_{u}; 
        \textbf{p}_{u}^{S} = \beta_{(S,u)}\textbf{s}_{u} + \beta_{(C,u)}\textbf{c}_{u}
    \end{equation}
The generated two feature embeddings $\textbf{p}_{u}^{I}$ and $\textbf{p}_{u}^{S}$  
represent the user's preferences for items and other users after transferring the shared knowledge between the two domains.

For predicting the scores of each item and user, we adopt a neural form MF \cite{he2017neural} to utilize the user's feature embedding.
For each task, a specific output layer is employed. The scores of user $u$ for item $v$ 
are calculated as follow:
\begin{equation}
    \hat{r}_{uv} = \textbf{W}_{I} (\textbf{p}_{u}^{I} \odot \textbf{q}_{v}); 
    \hat{x}_{ut} = \textbf{W}_{S} (\textbf{p}_{u}^{S} \odot \textbf{g}_{t})
\end{equation}
% trustSVD, NCF, DiffNet
% Distinct from the most model in previous works \cite{he2017neural,wu2019neural} 
% which use Negative Sampling strategy to train the model, 
% eatnn
$\textbf{q}_v$ and $\textbf{g}_t$ denotes the latent factor vector 
of item $v$ and user $t$ as a friend, respectively.
The operation $\odot$ denotes the element-wise product of vectors

Whole-data based strategy leverages the full data with a potentially better coverage.
Thus we adopt an efficient whole-data train strategy \cite{chen2019efficient} to optimize our model.
For each task, the loss functions are defined as follow:
\begin{small}
    \begin{equation}
        \begin{aligned}
            \tilde{\mathcal{L}}_{I}(\Theta) & = \sum^{D_1}_{i=1}\sum^{D_1}_{j=1} 
                \bigg( (h_{I,i}h_{I,j}) \bigg( \sum_{u \in \mathcal{B}} p_{u,i}^{I}p_{u,j}^{I} \bigg) 
                \bigg( \sum_{v \in \mathcal{V}} c_{v}^{I-} q_{v,i}q_{v,j} \bigg) \bigg) \\
                & + \sum_{u \in \mathcal{B}} \sum_{v \in \mathcal{V}^{+}}\bigg( (1 - c_{v}^{I-}) \hat{r}_{uv}^{2} - 2\hat{r}_{uv}\bigg)
        \end{aligned}
    \end{equation}   
    \begin{equation}
        \begin{aligned}
            \tilde{\mathcal{L}}_{S}(\Theta) & = \sum^{D_1}_{i=1}\sum^{D_1}_{j=1} 
                \bigg( (h_{S,i}h_{S,j})  \bigg( \sum_{u \in \mathcal{B}} p_{u,i}^{S}p_{u,j}^{S} \bigg) 
                \bigg( \sum_{t \in \mathcal{U}} c_{t}^{S-} g_{t,i}g_{t,j} \bigg) \bigg) \\
                & + \sum_{u \in \mathcal{B}} \sum_{t \in \mathcal{U}^{+}}\bigg( (1 - c_{t}^{S-}) \hat{x}_{ut}^{2} - 2\hat{x}_{ut}\bigg)
        \end{aligned}
    \end{equation}     
\end{small}
$I$ and $S$ are related to the item domain and social domain.
$D_1$ is the latent factor number. The scalar $h$,$p$,$q$,$g$ denote
the element of their corresponding vectors \bm{$h$},\bm{$p$},\bm{$q$},\bm{$g$}.
$i$ and $j$ denote the index of element in the vector.
$U^+$ and $V^+$ denote the items $v$ have interacted and the friends
that directly connect. $B$ is batch of users. $c_v^{I-}$ and $c_t^{S-}$ are the weight of negative instances in two domains.

Both rating and social information are very sparse which could lead to the overfitting problem. 
We consider that there has an implicit correlation between common knowledge and domain-specific knowledge. 
This assumption motivates us to propose a novel regularization term to against the overfitting problem:
\begin{small}
    \begin{equation}
        \begin{aligned}
            \tilde{\mathcal{L}}_{Reg}(\Theta) = \sum_{k} (\| \textbf{i}^{k} - \theta_{\alpha}^{k} \textbf{c}^{k} \|^{2} + 
                \| \textbf{s}^{k} - \theta_{\beta}^{k} \textbf{c}^{k} \|^{2})
        \end{aligned}
    \end{equation}
\end{small}
Where $\theta$ represents the weight of common knowledge $\textbf{c}$.
$\alpha$ and $\beta$ are related to the item domain and social domain.

After that, we integrate both the sub-tasks loss and the novel regularization term into an overall objective function as follow:
\begin{equation}
    \mathcal{L}(\Theta) = \tilde{\mathcal{L}}_{I}(\Theta) + \lambda_1\tilde{\mathcal{L}}_{S}(\Theta) + \lambda_2\tilde{\mathcal{L}_{Reg}}(\Theta) + \lambda_3\| \Theta \|^{2}
\end{equation}
$\Theta$ represents the parameters of our model. $\lambda_1$,$\lambda_2$,and $\lambda_3$ are the parameters to adjust the weight proportion of each term.

\section{Experiments}

\begin{table*}[]
    \scriptsize
    \caption{Performance of all the comparison methods on the Ciao and Yelp datasets.
        The last column “Avg Imp" indicates the average improvement of PTLN over the corresponding baseline on average.
        N indicates top-N task.}
    \begin{center}
    \setlength{\tabcolsep}{0.2mm}{
        \begin{tabular}{@{}l|c|c|c|c|c|c|c|c|c|c|c|c|c}
            \hline
            \diagbox{Baselines}{Metrics}&
            \multicolumn{3}{|c|}{Precision}&
            \multicolumn{3}{c|}{Recall}&
            \multicolumn{3}{c|}{NDCG}&
            \multicolumn{3}{c|}{MRR} \\
            \hline
            \textbf{Ciao}   & \textbf{N=5}  & \textbf{N=10} & \textbf{N=15} & \textbf{N=5}  & \textbf{N=10} & \textbf{N=15} & \textbf{N=5}  & \textbf{N=10} & \textbf{N=15} & \textbf{N=5}  & \textbf{N=10} & \textbf{N=15} & \textbf{Avg Imp}       \\ \hline
            \textbf{BPR}    & 0.0208          & 0.017           & 0.0141          & 0.0272          & 0.0496          & 0.0631          & 0.0289          & 0.036           & 0.0402          & 0.0479          & 0.0538         & 0.056           & \textbf{60.84\%} \\ \hline
            \textbf{NCF}    & 0.0217          & 0.0176          & 0.0149          & 0.0392          & 0.057           & 0.0721          & 0.0294          & 0.0385          & 0.0441          & 0.0508          & 0.057          & 0.0596          & \textbf{45.92\%} \\ \hline
            \textbf{SAMN}   & 0.0266          & 0.0225          & 0.0195          & 0.0482          & 0.0743          & 0.0959          & 0.0405          & 0.0506          & 0.0575          & 0.0562          & 0.0632         & 0.0653          & \textbf{17.47\%} \\ \hline
            \textbf{EATNN}  & 0.0295          & 0.0233          & 0.0195          & 0.0528          & 0.0763          & 0.094           & 0.0454          & 0.054           & 0.0598          & 0.071           & 0.0787         & 0.0816          & \textbf{6.53\%} \\ \hline
            \textbf{PTLN} & \textbf{0.0307} & \textbf{0.0244} & \textbf{0.0203} & \textbf{0.0571} & \textbf{0.0818} & \textbf{0.1006} & \textbf{0.0494} & \textbf{0.0585} & \textbf{0.0646} & \textbf{0.0755} & \textbf{0.083} & \textbf{0.0866} &                 \\ \hline
        \end{tabular}
        \\
        \begin{tabular}{@{}l|c|c|c|c|c|c|c|c|c|c|c|c|c}
            \hline
            \diagbox{Baselines}{Metrics}&
            \multicolumn{3}{|c|}{Precision}&
            \multicolumn{3}{c|}{Recall}&
            \multicolumn{3}{c|}{NDCG}&
            \multicolumn{3}{c|}{MRR} \\
            \hline
            \textbf{Yelp}   & \textbf{N=5}  & \textbf{N=10} & \textbf{N=15} & \textbf{N=5}  & \textbf{N=10} & \textbf{N=15} & \textbf{N=5}  & \textbf{N=10} & \textbf{N=15} & \textbf{N=5}  & \textbf{N=10} & \textbf{N=15} & \textbf{Avg Imp}       \\ \hline
            \textbf{BPR}   & 0.0349          & 0.0282          & 0.0235          & 0.0317            & 0.0497             & 0.0601             & 0.0507          & 0.0537           & 0.0548           & 0.0832          & 0.092           & 0.0949          & \textbf{16.03\%}  \\ \hline
            \textbf{NCF}    & 0.0337          & 0.0292          & 0.0266          & 0.0503            & 0.0626             & 0.0711             & 0.0429          & 0.0465           & 0.0487           & 0.0721          & 0.081           & 0.0843          & \textbf{24.74\%}  \\ \hline
            \textbf{SAMN}   & 0.0333          & 0.0283          & 0.0276          & 0.0496            & 0.0568             & 0.0695             & 0.045           & 0.047            & 0.0511           & 0.0745          & 0.0822          & 0.0865          & \textbf{12.96\%}  \\ \hline
            \textbf{EATNN}  & 0.0327          & 0.029           & 0.0266          & 0.0507            & 0.0579             & 0.0666             & 0.0462          & 0.049            & 0.0516           & 0.0749          & 0.0835          & 0.087           & \textbf{12.18\%}  \\ \hline
            \textbf{PTLN} & \textbf{0.0356} & \textbf{0.0307} & \textbf{0.0274} & \textbf{0.0558}   & \textbf{0.0647}    & \textbf{0.0714}    & \textbf{0.0543} & \textbf{0.057}   & \textbf{0.0587}  & \textbf{0.0889} & \textbf{0.0978} & \textbf{0.1009} &                  \\ \hline
        \end{tabular}}

        \label{tab:all}
    \end{center}
\end{table*}

\subsection{Experimental Settings}

\subsubsection{Dataset} We experimented with two public datasets: \textbf{Ciao}\cite{chen2019efficient}
and \textbf{Yelp}\cite{shi2015semantic}.
Ciao provides a large amount of rating information and social information, while users make friends 
with others and express their experience through the form of reviews and ratings on Yelp. 
The two datasets were constructed following previous work \cite{chen2019efficient,shi2015semantic}. 
Each dataset contains users’ ratings of the items they 
have interacted with and the social connections between users. To address the Top-N recommendation task, we remove all ratings that less than 4 for all datasets and keep others with a score of 1. 
This preprocessing method aims at recommending the item list that users liked, 
and is widely used in existing works \cite{chen2019efficient,chen2019social,wu2016collaborative}. 

\textbf{Baselines} To evaluate the performance of Top-K recommendation,we compare our PTLN with the following methods: \textbf{BPR} \cite{rendle2012bpr}: A classic and widely used ranking algorithm for recommendation. 
    It is implemented by learning pairwise relation of rated and unrated items for each user rather than direct learning to predict ratings.
\textbf{NCF} \cite{he2017neural}: A neural CF model combines element-wise and 
    hidden layers of the concatenation of user and item embedding to capture their high-order interactions.
\textbf{SAMN} \cite{chen2019social}: A state-of-the-art deep learning method leverages attention mechanism to model both aspect- and friend-level differences 
    for the social-aware recommendation. \textbf{EATNN} \cite{chen2019efficient}: A state-of-the-art method uses attention mechanisms 
    to adaptively capture the interplay between item domain and social domain for each user.
% \subsubsection{Evalutation Metrics}

\textbf{Evalutation Metrics} We adapt four popular metrics Precision, Recall, 
NDCG( Normalized Discounted Cumulative Gain), and MRR(Mean Reciprocal Rank) for evaluation. 
Specifically, NDCG is a position-aware ranking metric, which assigns a higher score to 
hits at higher positions. MRR considers the ranking position of the first correct 
item in the recommended list. The higher value of these evaluation metrics, 
the better performance of the recommender system.

% \subsubsection{Parameter Setting}
\textbf{Parameter Setting}
The parameters for all baseline methods were initialized as in the corresponding 
papers and were then carefully tuned to achieve optimal performance.
The learning rate for all models were tuned among [0.0005, 0.0001, 0.005, 0.001, 0.05, 0.01].
To prevent overfitting, we tuned the dropout ratio in [0.5,0.7,0.9].
The batch size was tested in [16,32,64,128,256], the embedding size $D_1$ and the dimension of attention 
network $D_2$ were tested in [32,64,128,256].
For our PTLN model, $D_1$ and $D_2$ were set to 128 and 32 on Ciao and set to 64 and 32 on Yelp.
The learning rate was set to 0.0005 
when using the Yelp and 0.01 
when using Ciao. The dropout ratio $\rho$ was set to 0.7 on both datasets.

\subsection{Performance Comparison }
We investigate the Top-N performance with N set to [5,10,15],
according with the real recommendation scenario.
We observe the results in Table \ref{tab:all}:

% \section{Experiments}
\begin{figure*}[h]
    \centering
    \includegraphics[width=0.8\textwidth]{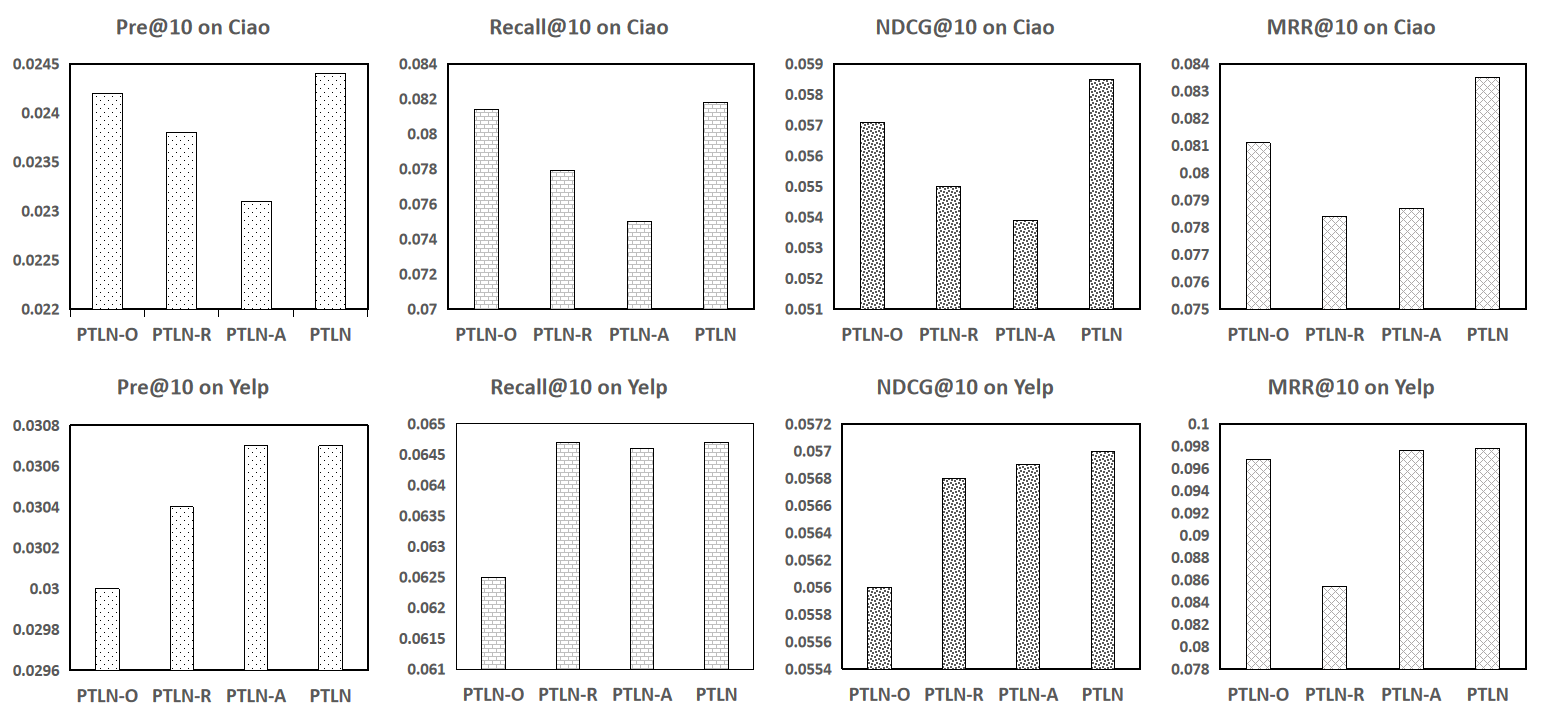}
    \caption{Performance with different PLTN variants on Ciao and Yelp.}
    \label{fig:ablation}
\end{figure*}

\begin{enumerate}
    \item Methods incorporating social information generally perform better than non-social method. SAMN, EATNN, and PTLN perform better than BPR and NCF. 
    This result is consistent with previous work which indicates that social information reflects users’ 
    interest and is helpful in the recommendation.
    \item Our method PTLN achieves the best performance on the two datasets and significantly 
    outperforms all baseline methods. Specifically, 
    compared to EATNN which is the best baseline that uses attention mechanisms to adaptively capture 
    the interplay between item domain and social domain for each user.  
    PTLN improves over EATNN about 6.53\% on Ciao and 12.18\% on Yelp. 
    The substantial improvement of our model over the baselines could be attributed to two reasons: 
    1) our model considers the propagation of social domain knowledge, item domain knowledge and common knowledge, 
    which allows the latent factor to be modeled with a finer granularity; 
    2) we consider the difference of friends' influence and the order bias.
\end{enumerate}
\begin{table}
    % \scriptsize    
    \caption{Performance with different propagation depth K on Ciao and Yelp.}

\centering
\resizebox{\textwidth}{!}{%
\begin{tabular}{|lllll|lllll|}
\hline
\textbf{Ciao} & Pre@10          & Recall@10       & NDCG@10         & MRR@10          & \textbf{Yelp} & Pre@10          & Recall@10       & NDCG@10        & MRR@10          \\ \hline
K=1           & 0.0242          & 0.0802          & 0.0559          & 0.0791          & K=1           & 0.0306          & 0.0619          & 0.0559         & 0.0966          \\ \hline
K=3           & 0.0242          & 0.0799          & 0.0557          & 0.0798          & K=3           & 0.0301          & 0.0617          & 0.0558         & 0.0973          \\ \hline
\textbf{K=2}  & \textbf{0.0244} & \textbf{0.0818} & \textbf{0.0585} & \textbf{0.0835} & \textbf{K=2}  & \textbf{0.0307} & \textbf{0.0647} & \textbf{0.057} & \textbf{0.0978} \\ \hline
\end{tabular}%
}

    \label{tab:order}
\end{table}
\textbf{Analyze of the propagation depth K} The number of propagation layers K reflects the extent to which the model uses social information and the degree to which social information influences the model. Table~\ref{tab:order} shows the results of different K values for both datasets.
When K increases from 1 to 2 ,the performance increases, while the performance drops when K=3. 
We empirically conclude that when the depth equals two  is enough for the social recommendation.

\subsection{Ablation Study}

\subsubsection{Impact of the order bias} A key characteristic of our proposed model is the order bias which 
considers the order of user’s friend in general. 
% As shown in the table 4, adding order embedding to our model leads to improvements, 
% which verifies our assumption that we should consider the extent of the influence 
% of users' different order friends.
PTLN-O denotes a variant model of PTLN without using order bias.
We can see that order bias has dramatically improved performance in Figure~\ref{fig:ablation}.
We speculate a possible reason is that order bias can dynamically 
adjust the output after fusing so that the updated user embedding of this order 
can better reflect the preference of the user after being influenced by friends.

\subsubsection{Impact of the attention mechanism} Another critical characteristic of our proposed model is we considering the diversity of friends' influence
by attention mechanism. 
% We show the results in Figure~\ref{fig:ablation}, with 
PTLN-A directly aggregate the friends' influence and user’s embedding without any attention learning process. 
From Figure~\ref{fig:ablation}, we can see that our model has a 
notable improvement in performance on the Ciao dataset when considering the difference of friends' influence.
However, results on the Yelp dataset is not as significant as Ciao. 
This observation implies that the usefulness of considering the importance strength of different elements in the modeling process varies, 
and our proposed friend-level attention modeling could adapt to different datasets’ requirements.

% \begin{table}[]
%     \scriptsize
%     \begin{tabular}{lllll|l}
%     \hline
%         \textbf{Ciao}      & \textbf{Pre@10} & \textbf{Recall@10} & \textbf{NDCG@10} & \textbf{MRR@10} & \textbf{Improve} \\ \hline
%         \textbf{PTLN-O}  & 0.0242          & 0.0814             & 0.0571           & 0.0811          & \textbf{1.68\%}  \\ \hline
%         \textbf{PTLN-R}  & 0.0238          & 0.0779             & 0.055            & 0.0784          & \textbf{5.1\%}   \\ \hline
%         \textbf{PTLN-FA} & 0.0231          & 0.075              & 0.0539           & 0.0787          & \textbf{7.33\%}  \\ \hline
%         \textbf{PTLN}    & \textbf{0.0244} & \textbf{0.0818}    & \textbf{0.0585}  & \textbf{0.0835} & \textbf{}        \\ \hline
%     \end{tabular}
%     \begin{tabular}{lllll|l}
%         \hline
%         \textbf{Yelp}      & \textbf{Pre@10} & \textbf{Recall@10} & \textbf{NDCG@10} & \textbf{MRR@10} & \textbf{Improve} \\ \hline
%         \textbf{PTLN-O}  & 0.03            & 0.0625             & 0.056            & 0.0968          & \textbf{2.17\%}  \\ \hline
%         \textbf{PTLN-R}  & 0.0304          & 0.0647             & 0.0568           & 0.0854          & \textbf{3.94\%}  \\ \hline
%         \textbf{PTLN-FA} & 0.0307          & 0.0646             & 0.0569           & 0.0976          & \textbf{0.13\%}  \\ \hline
%         \textbf{PTLN}    & \textbf{0.0307} & \textbf{0.0647}    & \textbf{0.057}   & \textbf{0.0978} &                  \\ \hline
%     \end{tabular}
%     \caption{Pre@10, Recall@10, NDCG@10 and MRR@10 performance with different PLTN variants}
% \end{table}

\subsubsection{Impact of the novel regularization} To evaluate the effectiveness of the proposed correlative regularization, 
we compare PTLN-R, a variant model of PTLN
 without using novel regularization, with PLTN, in Figure~\ref{fig:ablation}.
The PTLN model performs better than the PTLN-R, proving that our novel regularization can make the algorithm more stable.

\section{Conclusions}

In this paper, we present a novel social-aware recommendation model PTLN to 
address the sparsity problem of data. The core component of our model is propagation 
layers that learn user embedding of each order by leveraging high-order information 
from the social domain, item domain, and common knowledge between the two domains. 
Attention mechanism and the concept of order bias are further employed to better distinguish the 
influence of different user friends. 
The proposed PTLN consistently and 
significantly outperforms the state-of-the-art recommendation models on different evaluation metrics, 
especially on the dataset with complicated social relationships and fewer item interactions which verified 
our hypothesis about the varying degrees of different friends’ influence.


\begin{thebibliography}{8}
\bibitem{hu2008collaborative}
Yifan Hu, Yehuda Koren, and Chris Volinsky.
\newblock Collaborative filtering for implicit feedback datasets.
\newblock In {\em 2008 Eighth IEEE International Conference on Data Mining},
  pages 263--272. IEEE, 2008.

\bibitem{rendle2012bpr}
Steffen Rendle et~al.
\newblock Bpr: Bayesian personalized ranking from implicit feedback.
\newblock {\em arXiv preprint arXiv:1205.2618}, 2012.

\bibitem{xiao2017learning}
Lin Xiao  Zhang Min, Zhang Yongfeng, Liu Yiqun, and Shaoping Ma.
\newblock Learning and transferring social and item visibilities for
  personalized recommendation.
\newblock In {\em CIKM 2017}, pages 337--346, 2017.

\bibitem{chen2019efficient}
Chong Chen, Min Zhang, Chenyang Wang, Weizhi Ma, Minming Li, Yiqun Liu, and
  Shaoping Ma.
\newblock An efficient adaptive transfer neural network for social-aware
  recommendation.
\newblock In {\em SIGIR 2019}, pages 225--234, 2019.

\bibitem{zhao2014leveraging}
Tong Zhao et~al.
\newblock Leveraging social connections to improve personalized ranking for
  collaborative filtering.
\newblock In {\em Proceedings of the 23rd ACM international conference on
  conference on information and knowledge management}, pages 261--270, 2014.

\bibitem{guo2015trustsvd}
Guibing Guo et~al.
\newblock Trustsvd: Collaborative filtering with both the explicit and implicit
  influence of user trust and of item ratings.
\newblock In {\em Proceedings of the AAAI Conference on Artificial
  Intelligence}, volume~29, 2015.

\bibitem{koren2008factorization}
Yehuda Koren.
\newblock Factorization meets the neighborhood: a multifaceted collaborative
  filtering model.
\newblock In {\em Proceedings of the 14th ACM SIGKDD international conference
  on Knowledge discovery and data mining}, pages 426--434, 2008.

\bibitem{chen2019social}
Chong Chen et~al.
\newblock Social attentional memory network: Modeling aspect-and friend-level
  differences in recommendation.
\newblock In {\em WSDM 2019}, pages 177--185, 2019.

\bibitem{qiu2018deepinf}
Jiezhong Qiu et~al.
\newblock Deepinf: Social influence prediction with deep learning.
\newblock In {\em Proceedings of the 24th ACM SIGKDD International Conference
  on Knowledge Discovery \& Data Mining}, pages 2110--2119, 2018.

\bibitem{wu2019neural}
Le~Wu, Peijie Sun, Yanjie Fu, Richang Hong, Xiting Wang, and Meng Wang.
\newblock A neural influence diffusion model for social recommendation.
\newblock In {\em SIGIR 2019}

\bibitem{wu2020diffnet++}
Le~Wu et~al.
\newblock Diffnet++: A neural influence and interest diffusion network for
  social recommendation.
\newblock {\em arXiv preprint arXiv:2002.00844}, 2020.

\bibitem{eaton2011selective}
Eric Eaton et~al.
\newblock Selective transfer between learning tasks using task-based boosting.
\newblock In {\em Proceedings of the AAAI Conference on Artificial
  Intelligence}, 2011.

\bibitem{lu2013selective}
Zhongqi Lu et~al.
\newblock Selective transfer learning for cross domain recommendation.
\newblock In {\em Proceedings of the 2013 SIAM International Conference on Data
  Mining}, pages 641--649. SIAM, 2013.

\bibitem{cao2010adaptive}
Bin Cao, Sinno~Jialin Pan, Yu~Zhang, Dit-Yan Yeung, and Qiang Yang.
\newblock Adaptive transfer learning.
\newblock In {\em AAAI}, volume~2, page~7, 2010.

\bibitem{moon2017completely}
Seungwhan Moon and Jaime~G Carbonell.
\newblock Completely heterogeneous transfer learning with attention-what and
  what not to transfer.
\newblock In {\em IJCAI}, volume~1, 2017.

\bibitem{li2020ddtcdr}
Pan Li and Alexander Tuzhilin.
\newblock Ddtcdr: Deep dual transfer cross domain recommendation.
\newblock In {\em Proceedings of the 13th International Conference on Web
  Search and Data Mining}, pages 331--339, 2020.

\bibitem{he2017neural}
Xiangnan He et~al.
\newblock Neural collaborative filtering.
\newblock In {\em Proceedings of the 26th international conference on world
  wide web}, pages 173--182, 2017.

\bibitem{shi2015semantic}
Chuan Shi et~al.
\newblock Semantic path based personalized recommendation on weighted
  heterogeneous information networks.
\newblock In {\em CIKM 2015}, pages 453--462, 2015.

\bibitem{wu2016collaborative}
Yao Wu, Christopher DuBois, Alice~X Zheng, and Martin Ester.
\newblock Collaborative denoising auto-encoders for top-n recommender systems.
\newblock In {\em WSDM 2016}

\end{thebibliography}
\end{document}